# Hardware Implementation of Hyperbolic Tangent Function using Catmull-Rom Spline Interpolation

Mahesh Chandra
NXP Semiconductors
India
mahesh.chandra_1@nxp.com

*Abstract*—Deep neural networks yield the state of the art results in many computer vision and human machine interface tasks such as object recognition, speech recognition etc. Since, these networks are computationally expensive, customized accelerators are designed for achieving the required performance at lower cost and power. One of the key building blocks of these neural networks is non-linear activation function such as sigmoid, hyperbolic tangent (tanh), and ReLU. A low complexity accurate hardware implementation of the activation function is required to meet the performance and area targets of the neural network accelerators. This paper presents an implementation of tanh function using the Catmull-Rom spline interpolation. State of the art results are achieved using this method with comparatively smaller logic area.

## I. INTRODUCTION

Artificial neural networks (ANNs) have been used for modeling the complex non-linear relationships between the inputs and outputs in multiple applications. An ANN consists of a layered network of the artificial neurons which compute the weighted sum of multiple inputs and pass it through a non-linear activation function. State of the art deep neural networks (DNNs) have many such layers connected in feed forward fashion [1]. Till a few years back, sigmoid and hyperbolic tangent (tanh) were used more frequently to model the non-linear activation of artificial neurons. However, simpler activation functions such as rectified linear unit (ReLU) have been more popular in recent deep neural networks. Using these DNNs, state of the art result has been achieved in various applications such as object detection and classification. However, there is another set of applications which requires the neural networks to model the history or sequence such as the natural language processing, classification of video sequences, and image captioning etc. Recurrent neural networks (RNNs) and long short-term memory (LSTM) have been used for such applications [2]. These neural networks continue to use tanh activation function for its ability to handle vanishing gradients and ease of computing gradient.

Since, these algorithms require huge computing resources; there has been an effort to implement dedicated accelerators to speed up the execution. Activation function is one of the key building block required for the efficient hardware accelerator. Experimental study has shown that the accuracy of the activation function impacts the performance and the size of the neural networks [3]. Hyperbolic tangent function, being a non-linear, function requires specific consideration for the accuracy and area trade-off. In this paper, a method is presented which achieves very good accuracy with a relatively smaller logic.

## II. METHODS FOR IMPLEMENTING TANH FUNCTION IN HARDWARE

This section presents the state of the art methods for tanh implementation. Tanh function, shown in figure 1, is a non-linear function defined as:

$$tanh(x) = \frac{e^x - e^{-x}}{e^x + e^{-x}} \quad (1)$$

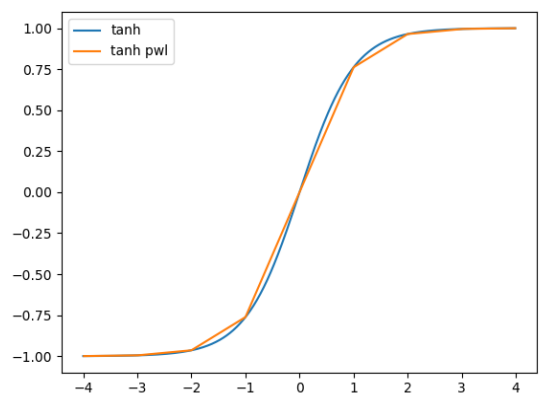

Fig. 1. tanh function and its piecewise linear approximation

The simplest implementation is to store the values of the function in a lookup table (LUT) and approximate the output with the lookup table value for the nearest input. Since, the function is non-uniform, it is difficult to balance the tradeoff between accuracy and area if the range is divided in equal sub-ranges. To address this issue, range addressable lookup table has been proposed by Leboeuf et al. [4]. The step size is varied depending on the variability of the function to reduce the size of LUT without impacting the accuracy. Another variation of this method is to use a two-step LUT. The first one with the coarse estimation and second one with the finer estimation. Namin et al. use this method but instead of an LUT, they use a combination of linear and saturation values for coarse approximation [5].

Zamanlooy et al. take advantage of the tanh function being an odd function and divide it in three ranges based on its basic properties; pass region, processing region and saturation region [6]. Then the hardware is optimized specific to the regions. In the pass region, the data is simply shifted and in the saturation region it is fixed. In processing region, data is mapped from the input by simple bit-level mapping (i.e. the combinatorial logic).

The function value can be interpolated by piecewise linear (PWL) approximation to reduce the error. The function value is stored in an LUT for known values and from these values, the function is interpolated for intermediate input values [7].

Adnan et al. have approximated the tanh function by Taylor series expansion [8]. The accuracy varies across the range of the input and the function is more accurately computed for smaller values of inputs. Moreover, if the number of terms in Taylor series are increased from three to four, improvement is just 2x where the error was large while it is 10x where the error was already small.

Gomar et al. approximate the tanh function by another simpler exponential function of base two for hardware implementation [9]. Their implementation requires an exponential unit, a division unit and supporting logic. RMSE error for this implementation is 0.0177 which is less than half of the range addressable LUT implementation.

Abdelsalam et al. have used the DCT (discrete cosine transform) interpolation filter (DCTIF) for tanh approximation [10]. Like [6], they also divide the tanh function in three regions and use DCTIF for approximation in processing region. This method achieves higher accuracy than any of the published methods. However, it requires huge memory for storing the coefficients. Thus, the existing methods either have low accuracy or high area and there is a need to balance the two.

### III. TANH INTERPOLATION USING CATMULL-ROM SPLINE

Splines are used for generating the curves of various shapes in the computer graphics applications. The approximation spline functions do not pass through the interpolating points while the interpolation spline function pass through them. Catmull-Rom spline function [11] is an interpolating spline, which has been used in many graphics and engineering applications. Given control points, the shape of the curve is fixed for the Catmull-Rom spline functions; however, there are variations of this function which can be adapted to a given shape [12, 13]. Moreover, a cubic Catmull-Rom spline function has only integer coefficients which reduces the implementation cost when compared to other spline functions. These properties of Catmull-Rom spline are very useful for the hardware implementation of tanh function.

The cubic Catmull-Rom spline is defined as:

$$f(x) = [t^3 \quad t^2 \quad t \quad 1] \begin{bmatrix} -1 & 3 & -3 & 1 \\ 2 & -5 & 4 & -1 \\ -1 & 0 & 1 & 0 \\ 0 & 2 & 0 & 0 \end{bmatrix} \begin{bmatrix} P_{k-1} \\ P_k \\ P_{k+1} \\ P_{k+2} \end{bmatrix} \quad (2)$$

Where,
'$P_i$' is value of function (e.g. tanh) at the uniformly sampled $x_i$ in given input range,
't' is between 0 and 1 and used to compute interpolation factor, and
f(x) is the interpolated value of the function at x for $x_k < x < x_{k+1}$

These equations can be used to interpolate tanh if we store the value of tanh for some points in the LUT. Table I and II show the RMS and maximum error analysis of tanh interpolation for different uniform sampling periods i.e. the difference between two adjacent $x_i$'s for which the function value is stored in LUTs. This analysis has been performed for 16-bit signed input x such that -4 < x < 4. This range is selected as the tanh function almost saturates beyond this range [6]. The precision for the input and output is considered equal as governed by input range. Note that for 16-bit input, 1 bit is used for sign and 2 bits for the integer part; so, there are 13-bits left for the fraction part. This fixes the precision to be to be $2^{-13}$ for both the input and output.

TABLE I. RMS ERROR FOR PWL AND CATMULL-ROM INTERPOLATION WITH DIFFERENT CONFIGURATIONS

| Sampling Period | LUT Depth | PWL | Catmull Rom | Accuracy Gain (x) |
|---|---|---|---|---|
| 0.5 | 8 | 0.008201 | 0.001462 | 5.61 |
| 0.25 | 16 | 0.002078 | 0.000147 | 14.16 |
| 0.125 | 32 | 0.000523 | 0.000052 | 10.02 |
| 0.0625 | 64 | 0.000135 | 0.000049 | 2.76 |

TABLE II. MAXIMUM ERROR FOR PWL AND CATMULL-ROM INTERPOLATION WITH DIFFERENT CONFIGURATIONS

| Sampling Period | LUT Depth | PWL | Catmull Rom | Accuracy Gain (x) |
|---|---|---|---|---|
| 0.5 | 8 | 0.023330 | 0.005179 | 4.50 |
| 0.25 | 16 | 0.006015 | 0.000602 | 9.99 |
| 0.125 | 32 | 0.001584 | 0.000152 | 10.42 |
| 0.0625 | 64 | 0.000470 | 0.000122 | 3.84 |

It can be observed from above tables that Catmull-Rom has better accuracy than the PWL interpolation. In the next section, we discuss the hardware implementation of the Catmull-Rom interpolation for tanh function.

### IV. DIGITAL CIRCUIT

This section describes the digital hardware implementation of the tanh function. For single bit RMS error, sampling period of 0.125 is good enough and hence, the implementation with 32-LUT is discussed here.

The cubic Catmull-Rom spline given in (2) can be re-written as:

$$f(x) = [P_{k-1} \quad P_k \quad P_{k+1} \quad P_{k+2}] \begin{bmatrix} -t^3 + 2t^2 - t \\ 3t^3 - 5t^2 + 2 \\ -3t^3 + 4t^2 + t \\ t^3 - t^2 \end{bmatrix} \quad (3)$$

The equation (3) can be considered as a dot product of two vectors and can be implemented by a simple MAC and vector computation units as shown in figure 2. The first vector, P vector, contains the control points, and the second vector, t vector, contains the interpolation factor.

The control points used in P-vector are stored in a look up table. Since tanh is an odd function, the size of control points LUT can be reduced by storing them only for the positive range. The LUT size is thus reduced and for a uniform interval of 0.125 and a range of x<4, only 32 values are required. The most significant five bits of the input are used as the index to the LUT. Since, the stored values are fixed (i.e. constant), we can use combinatorial logic instead of a memory cut to store these values. This LUT is a simple bit level mapping logic instead of the memory cut.

The second vector, t vector, contains the interpolation factor and can either be computed by a digital circuit or can be stored in the LUT depending on performance area tradeoff. The LUT implementation can be operated at higher frequency, while the polynomial computation logic is smaller in area. The digital logic needs to compute the values of the four cubic polynomials of t. As discussed in previous paragraph, msbs are used for addressing the LUT, the remaining bits (lsbs) can directly be used as t.

The precision for LUT entries is kept same as that of the input and output i.e. 13-bits. Figure 3 shows the final implementation with bit width at each level.

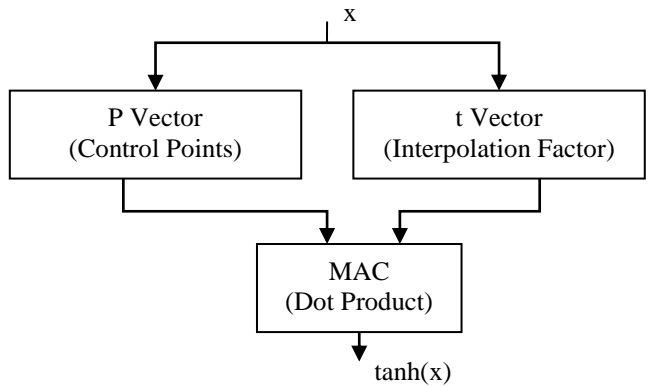

Fig. 2. High level block diagram of tanh function implementation

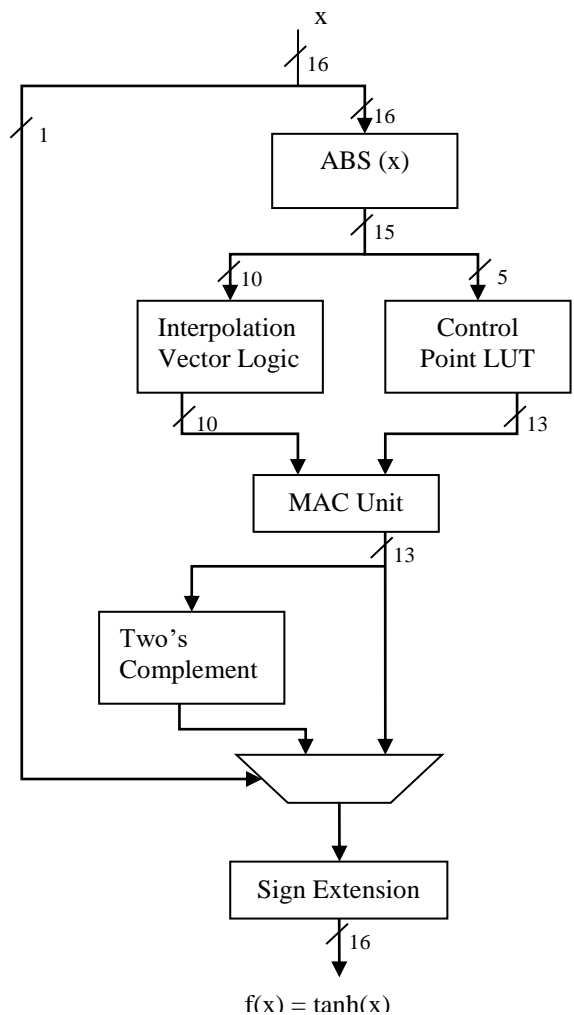

Fig. 3. Data Flow diagram for tanh function implementation

## V. RESULTS

The RTL code was written for the circuit and synthesized for comparative analysis. The circuit can be designed in multiple ways and different configurations were tried for area/timing trade-offs. It was observed that the circuit runs faster if the vector containing polynomial in 't' is also stored in LUTs; however, the area is larger in this case. The smallest area configuration has been used for comparison with the state of the art. This implementation has been synthesized for 500MHz clock frequency.

The comparison of accuracy and area with different implementations is given in the table III. As the results for [10] are published only for FPGA, the gate count is estimated by extrapolating it based on the published results. The DCTIF implementation [10] also requires a memory; the memory size has been considered as published.

TABLE III. AREA AND ACCURACY COMPARISON

| Work | Method | Precision (bits) | Gate Count and Memory | Accuracy |
|---|---|---|---|---|
| [5] | RALUT | 10 | 515 Gates; No Memory | 0.0189 |
| [6] | Region based processing | 6 | 129 Gates, No Memory | 0.0196 |
| [10] | DCTIF | 11 | 230 Gates + 22.17 Kbits Memory | 0.00050 |
| [10] | DCTIF | 16 | 800 Gates + 1250.5 Kbits Memory | 0.00010 |
| This | CR Spline | 13 | 5840 Gates, No Memory | 0.000152 |

It is obvious that for the proposed implementation is either significantly more accurate [5, 6, 10] or significantly smaller [10] than the other implementations. Note that the logic area of [10] is smaller; however, it also requires a huge memory.

## VI. CONCLUSION

This paper presented an implementation of tanh function using Catmull-Rom spline interpolation. This spline is used in many other engineering applications. It is shown that this approach yields state of the art accuracy at a significantly lower area. The implementation proposed here can be used in the hardware accelerators for the neural networks.